\documentclass[aps,12pt, a4paper]{revtex4}

\usepackage[utf8]{inputenc}
\usepackage[T1]{fontenc}
\usepackage{amsmath}
\usepackage{amsfonts}
\usepackage{amssymb}
\usepackage{graphics,graphicx}

\usepackage{graphicx}
\usepackage{subfigure} 
\usepackage{wrapfig} 
\usepackage{epstopdf}
\usepackage{xcolor}
\graphicspath{{figuras/}}

\usepackage[breaklinks=true]{hyperref}
\usepackage{setspace}
\usepackage{graphicx}
\usepackage{color}

\usepackage{tikz,tikz-3dplot}
\tdplotsetmaincoords{80}{45}

\tikzset{surface/.style={draw=black, fill=white, fill opacity=.6}}

\usepackage{tikz}
\usetikzlibrary{decorations.pathmorphing,patterns}

\begin{document}

\title{$q$-Deformed glueballs spectrum in AdS/QCD correspondence}

\author{Fabiano F. Santos $^{1}$ and F. A. Brito$^{2,3}$}
\email{fabiano.ffs23@gmail.com, fabrito@df.ufcg.edu.br}
\affiliation{$^1${Departamento de Física, Universidade Federal do Maranhão, Campus Universitario do Bacanga, São Luís (MA), 65080-805, Brazil.}\\
$^2${Departamento de F\'\i sica, Universidade Federal da Para\'iba, Caixa Postal 5008, 58051-970 Jo\~ao Pessoa PB, Brazil.}\\
$^3${Departamento de F\'\i sica, Universidade Federal de Campina Grande, Caixa Postal 10071, 58109-970, Campina Grande PB, Brazil.} }
\date{\today}

\begin{abstract}
 This work presents the application of the $q$-algebra in the glueballs spectrum. This algebra is implemented through Jackson derivatives in a Schrödinger-like equation resulting from the gravity fluctuations around the braneworld scenario in five dimensions. In our prescription, we consider a four-dimensional AdS$_4$ brane, living in AdS$_5$ bulk, that is also known for describing locally localized gravity via quasi-zero mode. At the appropriate limit, this background leads to confinement and allows us to find the shape of the $q$-deformed glueball spectrum in the AdS$_{5}$/QCD$_{4}$ correspondence. The introduction of the $q$-deformation provides us with a richer glueball spectrum.

\end{abstract}

\maketitle

\newpage
\section{Introduction}

The investigation into the braneworld spectrum has garnered significant attention in the scenario of the Einstein gravity \cite{Karch:2000ct,Csaki:2000fc,DeWolfe:1999cp,Nozari:2012pq,Nozari:2013ira,Nozari:2007qg} and beyond \cite{Brito:2018pwe,Fu:2019xtx,Santos:2021guj,Santos:2023eqp,Santos:2022uxm}. This scheme aims to extract the braneworld spectrum through gravity fluctuations. The problem of gravity localization in $AdS_{4}$ branes was first approached by Karch and Randall \cite{Karch:2000ct}, where an almost massless mode, i.e. the ``quasi zero'' mode overcomes the Kaluza-Klein modes \cite{Randall:1999vf,Randall:1999ee} and develops the locally localized gravity on the braneworld. This mechanism is more general in localizing gravity and requires no conditions in the space far from the brane. However, a mechanism of localizing gravity as in \cite{Csaki:2000fc} realizes the study about Minkowsky (M$_{4})$ thick branes. The scalar fields model to address thick branes (e.g., M$_{4}$, dS$_{4}$ and AdS$_{4}$ thick branes) employs $bulk$ scalar fields \cite{Csaki:2000fc,DeWolfe:1999cp,Flanagan:2001dy,Kobayashi:2001jd}. Besides, analytical solutions in first-order formalism  have been analyzed in the model \cite{DeWolfe:1999cp}, but only for the case where the four-dimensional cosmological constant on the brane $\Lambda_{4d}\!=\!0$. This model is also addressed in the context of the locally localization of gravity, but solutions were found only at the thin wall limit. In addition to these studies presented for the gravity localization scenarios in Einstein gravity, one can also include gauge and tensor vector fields in thick branes by using modified gravity theories such as $f(T)$ and $f(T,B)$ \cite{Moreira:2024nyy,Belchior:2023gmr,Moreira:2023pes,Moreira:2023pkh,Moreira:2023byr}. In these cases, the modified couplings produce a normalizable zero mode. Furthermore, another advantage of using these modified gravity models is the probing of resonant modes, where it is possible to understand the enormous spectrum of fields living on them.

In this work we shall consider one of the aforementioned gravity localization scenarios to serve as the gravity side of a particular ``deformed'' AdS/CFT correspondence known as AdS/QCD correspondence to address the glueball spectrum due to gravity fluctuations governed by a Schrödinger-like equation. In addition, in order to enrich the model to address more realistic phenomenology we propose the introduction of $q$-algebra in the context of locally localization of gravity by using the $q$-derivative operator well-known as Jackson derivative $(JD)$ \cite{Marinho:2011td,LARSSON1,Chaichian,Santos:2022fbq,Brito:2018wyw,Kanakoglou:2007vf}, where the $q$-derivative produces a generalization of the Schrödinger-like equation and its spectrum. This scenario is equipped with kinematic scalar field terms and scalar potential, which develops a non-zero kink profile in the bulk \cite{Gubser:2008px,Gubser:1996de,Burgess:1999vb,Policastro:2001yc,Policastro:2002se,Santos:2021orr,Santos:2023flb,Santos:2022lxj,Santos:2023mee}. These ingredients are fundamental for breaking the conformal symmetry and accessing AdS$_{5}$/QCD$_{4}$ correspondence, such that we can study, e.g., the linear confinement. In the Anti-de Sitter/Conformal Field Theory (AdS/CFT) and Anti-de Sitter/Quantum Chromodynamics (AdS/QCD) correspondence \cite{Brodsky:2011sk,Maldacena:1998im}, one uses $(d+1)$-dimensional AdS or modified background, respectively, to establish a correspondence with a quantum field theory in $d$-dimensions. An important implication of these dualities is the description of strong interactions. The phenomenological approximation known as AdS/QCD starts with a field theory in five dimensions motivated by string theory and QCD. In this description, the asymptotic linearity of Reggie's trajectories arises from some backgrounds. Such backgrounds are reduced to standard AdS space in the ultraviolet (UV) regime, whereas present modifications in infrared (IR) regime that turns out to be crucial to describe linear confining in QCD. Thus, it is natural to expect that interactions between quarks dominate in the Coulomb term at short distances and the linear term dominates over large distances \cite{Andreev:2006ct}. In this work, we investigate glueball trajectories \cite{Brower:2000rp,Brower:1999nj}, where we consider a $warp$ Gaussian factor in the Schrödinger-like equation; this is implied in a generalization of the masses. 

The idea behind our proposal is based on the study of quantum groups and quantum algebras that sparked interest in recent years. This has a large spectrum of applications such as cosmic strings, black holes, fractional quantum Hall effect, high-temperature superconductors, rational field theory, non-commutative geometry and quantum theory of super-algebras \cite{Chaichian}.  Furthermore, solutions for deformed Einstein equations and quantum black holes were considered. If the black hole is charged, the deformed solution imply that the deformation parameters lead the charged black hole to have a smaller mass than the usual Reissner-Nosdström black hole. In this sense, the typical black hole mass reduction is considered as a transition from a classical to quantum regime of the black hole \cite{Dil:2016vhx,Boumali:2016lns}. Beyond these applications, studies in dark matter have been considered in \cite{Dil:2016rmd}, where the coupling of this model assumes that the scalar plays the role of the {\it basic number}. These $q$-scalar fields modify the expansion of the Universe, where inhomogeneities provide soliton type attractor, which implies that the coupled $q$-deformed dark energy model is consistent with the conventional dark energy models, satisfying the Universe accelerated phase. In our case, we have that the fluctuations for the metric are deformed when we consider $q$-fluctuations. In this sense, in our description, these fluctuations are identified with the basic number. 

The paper is organized as follows: In Sec.~\ref{v0}, we present the scenario of five-dimensional gravity coupled to a scalar field \cite{Bazeia:2004yw} and find the Schrödinger-like equation through the fluctuations process \cite{Bazeia:2004yw,Brito:2018pwe,Brito:2019ose,DeWolfe:1999cp}. In Sec.~\ref{v}, we apply the Jackson derivative as presented by \cite{Marinho:2011td} into the Schrödinger-like equation for the gravity fluctuations. In Sec.~\ref{v1}, we use the harmonic oscillator-like $warp$ $factor$ and explore the AdS/QCD scenario. Finally, in Sec.~\ref{v2}, we make our conclusions about the implications of the $q$-algebra in our analysis.

\section{Gravity fluctuations}\label{v0}

In five-dimensions the gravity coupled to a scalar field \cite{Karch:2000ct,Csaki:2000fc,DeWolfe:1999cp,Randall:1999vf,Randall:1999ee,Bazeia:2004yw} is described by the following action
\begin{eqnarray}
S=\int d^{5}x\sqrt{g}\left[-\frac{1}{4}R+\frac{1}{2}\partial_{M}\phi \partial^{M}\phi-U(\phi)\right],\label{2.3}
\end{eqnarray}
where the signature is $(+----)$, $M=0,1,2,3,5,$ and $g=det(g_{MN})$. The metric Ansatz for this configuration can be written as
\begin{eqnarray}
ds^{2}=e^{2A(r)}\bar{g}_{ij}dx^{i}dx^{j}-dr^{2}.\label{2.4}
\end{eqnarray}
The linearization of the Einstein equations by considering the following perturbations $\bar{g}_{MN}=g_{MN}+h_{MN}$ with $U(\phi)=\Lambda$ \cite{Bazeia:2004yw,Brito:2018pwe,Brito:2019ose,DeWolfe:1999cp}, provides
\begin{eqnarray}
[\partial^{2}_{r}+4A'(r)\partial_{r}-e^{-2A(r)}(\Box_{4d}+2\Lambda_{4d})]\Phi_{ij}=0.\label{2.13}
\end{eqnarray}
Here $\Phi_{ij}$ describes the wave function of the graviton \cite{Bazeia:2004yw,Brito:2018pwe}. Now considering $\Phi_{ij}=h(r)\Omega(x^{\mu})$, where the equation describing the 4d graviton is $(\Box_{4d}+2\Lambda_{4d})\Omega=m^{2}\Omega$ with $h(r)=e^{3A(z)/2}\psi(z)$. By considering $z(r)=\int{e^{-A(r)}}dr$ into equation (\ref{2.13}), we have that 
\begin{eqnarray}
-\partial^{2}_{z}\psi(z)+V(z)\psi(z)=m^{2}\psi(z),\label{3.1} 
\end{eqnarray}
where $V(z)$ is given by
\begin{eqnarray}
V(z)=\frac{3}{2}A''(z)+\frac{9}{4}A'^{2}(z).\label{3.2} 
\end{eqnarray}
According to the $AdS_{5}/QCD_{4}$ correspondence, the equation (\ref{3.1}) that describes graviton in the bulk can describe particles with $J^{PC}=2^{++}$ such as glueballs on the $AdS_5$ boundary.  Here, scalar perturbations do not contribute directly, because their contributions are absorbed in the simplification of the equation (\ref{3.1}) as a consequence of the equations of motion at zeroth order. For more details about this procedure, see \cite{DeWolfe:1999cp}. Moreover, if we analyze the deduction of the equation (\ref{3.1}) more carefully, we can see that during the process presented by \cite{Bazeia:2004yw,Brito:2018pwe,DeWolfe:1999cp} due to the relation between $A(r)$ and $\phi(r)$ that comes from Einstein equation in a combination of the form $G_{00}+G_{55}=T_{00}+T_{55}$ with $8\pi\,G=1$ \cite{Bazeia:2004yw,DeWolfe:1999cp}, this combination helps us to simplify the equation of the fluctuations to the form presented in  (\ref{2.13}). 

\section{$q$-Deformation}\label{v}

In this section, we shall discuss the generalization of gravity fluctuations equation \eqref{3.1}, obtained in the braneworld scenario, by introducing $q$-deformed Heisenberg algebra in terms of creation and annihilation operators and $N$
\begin{equation}\label{q-algebra}
    [c,c]_k=[c^\dagger,c^\dagger]_k=0, \qquad c c^\dagger - k q c^\dagger c = q^{-N}, \quad [N,c^\dagger]=c^\dagger, \qquad [N,c]=-c,
\end{equation}
where $q$ is a real deformation parameter and $k=\pm 1$ stands for bosonic ($+$) and fermionic ($-$) $q$-algebras. In the Bargmann holomorphic representation $c=z$ and $c^\dagger=D^{(q)}_z$, where the latter  
is the Jackson derivative \cite{Marinho:2011td,LARSSON1}~\footnote{This should not be confused with fractional derivative that is defined as $D^\zeta(f)=d^{\zeta}f/dz^{\zeta}$, where $\zeta$ $\in \mathbb{R}{}$.}. In doing this,  we replace the ordinary differential operator to apply Jackson differential operator. The Jackson derivative \cite{Marinho:2011td,LARSSON1} is given by
\begin{eqnarray}\label{JD}
D^{(q)}_{z}h(z)&=&\frac{h(qz)-h(z)}{z(q-1)}.   
\end{eqnarray}
One can make transparent how this operator works by using the following simple example, without loss of generality, i.e., $D_x^{(q)} (x^\alpha)=[\alpha]_q\, x^{\alpha-1}$, where $[\alpha]_q$ is the {\it basic number} defined as
\begin{eqnarray}
[\alpha]_{q}=\frac{q^{\alpha}-1}{q-1}.\label{3.4}
\end{eqnarray}
By considering the Jackson derivative we can either use its exact form \eqref{JD} or approximations. As we know the wave-function $\psi$ that describes the gravity fluctuations, by construction, was assumed to be small, whereas the warp factor $A(z)$ was not restricted to this condition. Thus, we shall apply the exact form \eqref{JD} to replace ordinary derivatives of the potential \eqref{3.2} and the approximated form to the wave-function. Let us now address the later case. Performing a Taylor expansion of the basic number, we have
\begin{eqnarray}
\frac{q^{\alpha}-1}{q-1}\approx \frac{\ln q\, \alpha}{q-1}+\frac{(\ln q)^{2}\alpha^{2}}{2(q-1)}+\frac{(\ln q)^{3}\alpha^{3}}{6(q-1)}+{\cal O}(\alpha^4).
\end{eqnarray} 
At this limit we can now establish a relationship between Jackson and ordinary differential operators. This was first noticed in \cite{Marinho:2011td} on the context of $q$-deformed statistical mechanics in the limit of high temperatures. Thus we can approach the Jackson derivative as follows
\begin{eqnarray}\label{3.3}
D^{(q)}_{z}[\alpha]_{q}=\partial_{z}\alpha,
\end{eqnarray}
such that at first order we have
\begin{eqnarray}\label{3.5}
D^{(q)}_{z}=\frac{q-1}{\ln q}\partial_{z}\,.
\end{eqnarray}
Thus, in discussing the application of the equation (\ref{3.5}) within the context of \cite{Karch:2000ct,Bazeia:2004yw,Brito:2018pwe} we begin with the Schrödinger-like equation (\ref{3.1}), which takes the form: 
\begin{eqnarray}
-\left(\frac{q-1}{\ln(q)}\right)^{2}\partial^{2}_{z}\psi(z)+V_{q}(z)\psi(z)=m^{2}\psi(z),\label{3.6}
\end{eqnarray}
where the $V_{q}$ is the deformed potential, given as  
\begin{eqnarray}
V_{q}(z)=\frac{3}{2}A_{q}''(z)+\frac{9}{4}A_{q}'^{2}(z).\label{3.7}
\end{eqnarray} 
The $q$-derivatives for this potential in the exact form \eqref{JD} are:

\begin{eqnarray}
A'_{q}(z)&=&\frac{[\ln(qy(z))-\ln(y(z))]\sqrt{1-y^{2}(z)}}{y(z)(q-1)}\label{3.8}\\
A''_{q}(z)&=&\frac{\frac{[\ln(q^{2}y(z))-\ln(qy(z))]\sqrt{1-q^{2}y^{2}(z)}}{y(z)q(q-1)}}{y(z)(q-1)}\nonumber\\
         &-&\frac{\frac{[\ln(qy(z))-\ln(y(z))]\sqrt{1-y^{2}(z)}}{y(z)(q-1)}}{y(z)(q-1)},\label{3.9}
\end{eqnarray}
where $y(z)$ depends on the braneworld solution, as we shall see in the next section.
As we previously anticipated, the equation (\ref{3.6}) can be used to study the linear confinement, such as in QCD, which involves the square of the masses of the mesons with spin eigenvalues $S$ or radial excitation of the mesons for linear growth with $S$ and $n$ as proposed in \cite{Karch:2006pv}. This behavior can be reproduced within a 5-dimensional gravity  theory by assuming a holographic duality for QCD, namely, AdS$_{5}$/QCD$_{4}$ \cite{Brodsky:2011sk,Andreev:2006ct}.

\section{$q$-glueballs spectrum}\label{v1}

Now using the Schrödinger-like equation (\ref{3.6}) to find the excitation of the \textit{background} solution of the gravitational sector from \eqref{2.3} in the thin-wall limit, that coincides with the solutions of \cite{Karch:2000ct}, for example, braneworlds with negative cosmological constant, given by $AdS_4$ brane solutions, we have $y_{AdS_{4}}(z)=\sin[\sqrt{-\Lambda_{4d}}(z-z_{0})]$ and from Eqs.~(\ref{3.8})-(\ref{3.9}) the $q$-deformed potential reads 
\begin{eqnarray}
V_{q}(z)\!=\!\frac{3}{2\beta_{q}}\!\cot^{2}[\sqrt{-\Lambda_{4d}}(z\!-\!z_{0})]\!+\!\frac{3}{4\beta_{q}}\!\cot^{2}[\sqrt{-\Lambda_{4d}}(z\!-\!z_{0})]\sqrt{1\!-\!q^{2}\sin^{2}[\sqrt{-\Lambda_{4d}}(z\!-\!z_{0})]},\label{4.0}
\end{eqnarray}
that by considering $\sqrt{-\Lambda_{{4d}}}(z-z_0)\ll1$, we have
\begin{eqnarray}
-\partial^{2}_{z}\psi(z)+\left(a(z-z_{0})^{2}+\frac{b}{(z-z_{0})^{2}}\right)\psi(z)=E\psi(z).
\end{eqnarray}
Notice that from now on, for the sake of clarity and simplicity, we shall assume the statement $\Lambda_{4d}\equiv \Lambda$.
Thus, we have here $b=9/(4(-\Lambda)\beta^{2}_{q})$, $a=3q^{2}(-\Lambda)/(8\beta^{2}_{q})$ with $\beta_{q}=\left(\frac{q-1}{\ln(q)}\right)^{2}$ and $E=m^{2}/\beta_{q}$ and this potential is similar to the case of \cite{Karch:2006pv}. The eigenfunctions are given in terms of Legendre's polynomials
\begin{eqnarray}
\psi(z)=2^{(2+\sqrt{1+4b})/4}e^{-\sqrt{a}(z-z_{0})^{2}/2}(z-z_{0})^{(3+\sqrt{1+4b})/2}L^{\frac{1}{2} \left(\sqrt{1+4b}+2\right)-1}_{-\frac{a\sqrt{1+4b}+2a-E\sqrt{a}}{4a}}\left(\sqrt{a}(z-z_{0})^2\right).
\end{eqnarray}
For $-\frac{a\sqrt{1+4b}+2a-E\sqrt{a}}{4a}=\frac{n}{2}$, we have
\begin{eqnarray}
m^{2}_{n}=2\beta_{q}\sqrt{a}(n+1)+\beta_{q}\sqrt{a}\sqrt{1+4b}.\label{4.1}
\end{eqnarray}

In the holographic perspective, this equation can be understood as the glueballs mass spectrum \cite{Brower:2000rp}, where the states for $J^{PC}=2^{++}$ are presented in the Table \ref{tb}. Performing a comparison with glueball trajectories, we have
\begin{eqnarray}
m_{n}^{2}=\mu^{2}_{4}n^{2}+\mu^{2}_{4}\delta n+\mu^{2}_{4}\gamma.\label{4.2}
\end{eqnarray}
From the equations (\ref{4.1}) and (\ref{4.2}), we conclude that: 
\begin{eqnarray}
\mu^{2}_{4}\delta&=&2\beta_{q}\sqrt{a},\\
\mu^{2}_{4}\gamma&=&2\beta_{q}\sqrt{a}+\beta_{q}\sqrt{a}\sqrt{1+4b},
\end{eqnarray}
where $\mu^{2}_{4}=36\pi\left(\frac{\Gamma(2/3)}{\Gamma(1/6)}\right)^{2}$. For the two initial states, we can restrict the space of parameters as follows
\begin{eqnarray}
m_{0}^{2}&=&\mu^{2}_{4}\gamma=22.097,\\ 
m_{1}^{2}&=&\mu^{2}_{4}+\mu^{2}_{4}\delta+\mu^{2}_{4}\gamma=55.584.
\end{eqnarray}
Solving the system, we find the values $\delta=4.00319$ and $\gamma=0.824703$, 
where $\Lambda=-2(\mu^{2}_{4}\delta)^{2}/(3q^{2})$. With this, we have $q^{2}\Lambda=-478.609$, and this relation provides the value of $q=48.92$ with $\Lambda=-0.2$. We can see that the conditions on the system provide a specific value of $q$. The number $q$ is real and arbitrary, assuming the values $0<q<\infty$. In the symmetric formulation, we have the regions of validity $0<q<1$ or $1<q<\infty$. Under this consideration, some cases can be studied: in the limit as $q\rightarrow 1$, the basic numbers defined in \eqref{3.4} become i) $[\alpha]\rightarrow\alpha$ and ii) $[0]=0$ and $[1]=1$, iii) $[-\alpha]=-[\alpha]$ that is invariant under symmetric transformation $q\leftrightarrow q^{-1}$ \cite{LARSSON1,Dil:2016vhx,Boumali:2016lns,Dil:2016rmd,Marinho:2011td,Kanakoglou:2007vf}. 

\begin{table}[!ht]
\begin{center}
\begin{tabular}{|c|c|} \hline
$JPC\quad equation:$&$Prediction\quad 2^{++}:$\\ \hline
$n=0$&$22.097$\\ \hline
$n=1$&$55.584$\\ \hline
$n=2$&$102.456$\\ \hline
$n=3$&$162.722$\\ \hline
$n=4$&$236.400$\\ \hline
\end{tabular}
\caption{The table shows our predictions compared with the Glueballs results of \cite{Brower:2000rp} for the values $q=48.92$ and $\Lambda=-0.2$.}\label{tb}
\label{Tb1}
\end{center}
\end{table}

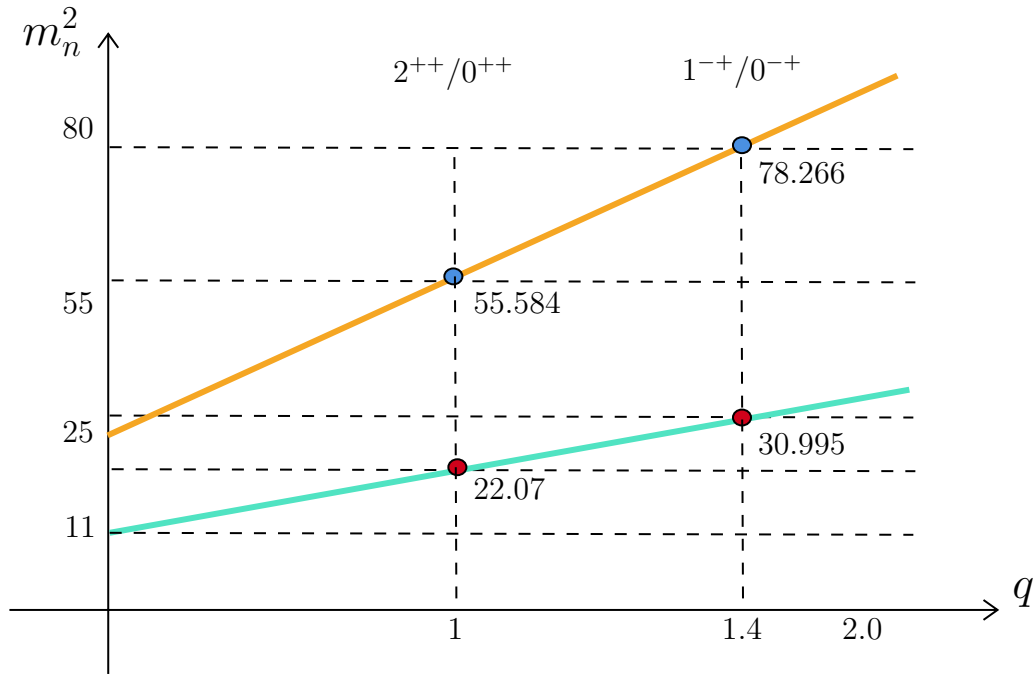
\begin{figure}
\centering
\tikzset{every picture/.style={line width=0.75pt}} 

\begin{tikzpicture}[x=0.75pt,y=0.75pt,yscale=-1,xscale=1]

\draw  (48,378.8) -- (541,378.8)(97.3,89) -- (97.3,411) (534,373.8) -- (541,378.8) -- (534,383.8) (92.3,96) -- (97.3,89) -- (102.3,96)  ;
\draw  [dash pattern={on 4.5pt off 4.5pt}]  (98,146) -- (499,147) ;
\draw  [dash pattern={on 4.5pt off 4.5pt}]  (99,213) -- (500,214) ;
\draw  [dash pattern={on 4.5pt off 4.5pt}]  (98,281) -- (499,282) ;
\draw  [dash pattern={on 4.5pt off 4.5pt}]  (99,308) -- (500,309) ;
\draw [color={rgb, 255:red, 80; green, 227; blue, 194 }  ,draw opacity=1 ][line width=2.25]    (98,340) -- (497,268) ;
\draw [color={rgb, 255:red, 245; green, 166; blue, 35 }  ,draw opacity=1 ][line width=2.25]    (97,291) -- (491,110) ;
\draw  [fill={rgb, 255:red, 74; green, 144; blue, 226 }  ,fill opacity=1 ] (265,211) .. controls (265,208.79) and (267.01,207) .. (269.5,207) .. controls (271.99,207) and (274,208.79) .. (274,211) .. controls (274,213.21) and (271.99,215) .. (269.5,215) .. controls (267.01,215) and (265,213.21) .. (265,211) -- cycle ;
\draw  [fill={rgb, 255:red, 74; green, 144; blue, 226 }  ,fill opacity=1 ] (409,145) .. controls (409,142.79) and (411.01,141) .. (413.5,141) .. controls (415.99,141) and (418,142.79) .. (418,145) .. controls (418,147.21) and (415.99,149) .. (413.5,149) .. controls (411.01,149) and (409,147.21) .. (409,145) -- cycle ;
\draw  [fill={rgb, 255:red, 208; green, 2; blue, 27 }  ,fill opacity=1 ] (409,282) .. controls (409,279.79) and (411.01,278) .. (413.5,278) .. controls (415.99,278) and (418,279.79) .. (418,282) .. controls (418,284.21) and (415.99,286) .. (413.5,286) .. controls (411.01,286) and (409,284.21) .. (409,282) -- cycle ;
\draw  [fill={rgb, 255:red, 208; green, 2; blue, 27 }  ,fill opacity=1 ] (267,307) .. controls (267,304.79) and (269.01,303) .. (271.5,303) .. controls (273.99,303) and (276,304.79) .. (276,307) .. controls (276,309.21) and (273.99,311) .. (271.5,311) .. controls (269.01,311) and (267,309.21) .. (267,307) -- cycle ;
\draw  [dash pattern={on 4.5pt off 4.5pt}]  (271,373) -- (270,148) ;
\draw  [dash pattern={on 4.5pt off 4.5pt}]  (414,373) -- (413,148) ;
\draw  [dash pattern={on 4.5pt off 4.5pt}]  (98,340) -- (499,341) ;

\draw (73,129.4) node [anchor=north west][inner sep=0.75pt]    {$80$};
\draw (462,381.4) node [anchor=north west][inner sep=0.75pt]    {$2.0$};
\draw (402,381.4) node [anchor=north west][inner sep=0.75pt]    {$1.4$};
\draw (265,381.4) node [anchor=north west][inner sep=0.75pt]    {$1$};
\draw (278,311.4) node [anchor=north west][inner sep=0.75pt]    {$22.07$};
\draw (420,288.4) node [anchor=north west][inner sep=0.75pt]    {$30.995$};
\draw (420,152.4) node [anchor=north west][inner sep=0.75pt]    {$78.266$};
\draw (278,217.4) node [anchor=north west][inner sep=0.75pt]    {$55.584$};
\draw (238,99.4) node [anchor=north west][inner sep=0.75pt]    {$2^{++} /0^{++}$};
\draw (382,98.4) node [anchor=north west][inner sep=0.75pt]    {$1^{-+} /0^{-+}$};
\draw (53,74.4) node [anchor=north west][inner sep=0.75pt]  [font=\Large]  {$m_{n}^{2}$};
\draw (547,361.4) node [anchor=north west][inner sep=0.75pt]  [font=\Large]  {$q$};
\draw (74,330.4) node [anchor=north west][inner sep=0.75pt]    {$11$};
\draw (73,282.4) node [anchor=north west][inner sep=0.75pt]    {$25$};
\draw (73,217.4) node [anchor=north west][inner sep=0.75pt]    {$55$};

\end{tikzpicture}
\caption{The squared masses $m_n^2$ as a function of $q$ for the modes $n=0,\,1$. At the undeformed limit $q=1$ we recover the glueballs spectrum $2^{++}/0^{++}$ as expected. For the deformed case $q\approx 1.41$ we emulate the $1^{-+}/0^{-+}$ sector. The deformed case $q=0.2294$ (not shown) leads to $m_0^2=5.6168$ or $m_0=2.3699$ GeV (or 2370 MeV) that approaches the resonance X(2370) recently pointed as a possible detection of the lightest pseudoscalar glueball with spin parity $0^{-+}$ \cite{BESIII:2023wfi}.}
\label{fig1}
\end{figure}

Alternatively, by using \eqref{4.1} evaluating the modes $n=0,\,1$ such that one recovers the modes without deformation, that is, at $q=1$, we find respectively, $\Lambda\approx -141.71$ and $\Lambda\approx -327.77$. We can now have the curves for arbitrary $q$ for fixed $\Lambda$. They are shown in Fig.~\eqref{fig1}. Notice that for specific values of q-deformed cases one can emulate other spin parity sectors, other than $2^{++}$ (or $0^{++}$, since the scalar sector produces the same spectra \cite{Bazeia:2004yw}) previously obtained for undeformed case $q=1$. This effect is not happening by an accident. We have a clue why in our scenario the q-deformation can emulate descriptions for arbitrary spin, at least for high spin \cite{Karch:2006pv}. This is because Eq.~\eqref{4.1} can be recast in the same structure of squared masses for high spin in terms of a deformed equation as follows
\begin{equation}\label{eq-high-spin}
    m_{n_n,S_q}^2=\alpha (n_q+S_q),
\end{equation}
where $\alpha$ is the slope, $n_q$ and $S_q$ are the q-deformed mode and spin, respectively, that are  defined as
\begin{equation}
    \alpha=\sqrt{-\Lambda}\sqrt{\frac{3}{2}},\qquad n_q=n\, q, \qquad S_q=q+\frac{q}{2}\sqrt{1+\frac{9}{(-\Lambda)}\left(\frac{\ln{q}}{q-1}\right)^4}.
\end{equation}
We emphasize that Eq.~\eqref{eq-high-spin} works well only for high spin and high modes, that is, for large $n_q$ and $S_q$, which is ensured for large $q$. In spite of this limitation to make full correlation with our case that deals with low spin, the structure of our equation \eqref{4.1} furnishes a nice way to pursue future investigations to provide further clarifications of the q-deformation in holographic scenarios such as in AdS/QCD correspondence.

Comparing our results from the Table \ref{Tb1} with lattice glueball spectrum, we can see that our spectrum produces similar results to QCD$_{4}$. As a result, any reasonable expectation of a strong coupling approximation should not give quantitative results \cite{Brower:2000rp}. However, we have a remarkable correspondence of the overall mass structure and rotation between our Strong Coupling Glueball Spectrum and the Lattice in a weak coupling to QCD$_{4}$ \cite{Brower:1999nj}. Similarly, the IIA-type rotation structure of supergravity is similar to the low-mass glueball spin splitting. Here, we have that the correspondence is sufficient to suggest that the Maldacena duality~\cite{Maldacena:1998im} may be correct, and thus, efforts to go beyond the strong coupling are noteworthy.

\section{Conclusion}\label{v2}

In this work, we generalize the Schrödinger-like equation that governs gravity fluctuations by replacing the ordinary derivatives with the $q$-derivative (Jackson derivative) operator. In the holographic perspective, the Schrödinger-like equation becomes an interesting mechanism to study the glueballs spectrum, where the Gaussian $warp$ $factor$ provides a linear spectrum. By solving the generalized Schrödinger-like with suitable boundary conditions we are able to find the glueball spectrum according to Tab.~\ref{tb} that are possible candidates for the scenario $J^{PC}=2^{++}$. Despite the logarithmic correction, the $q$-algebra has no singularity for any specific value of $q$. In obtaining the constants, $\delta$, and $\gamma$, we have determined analytically the value of the mode $m^{2}_{0}$, that agrees with the variational value estimated in \cite{Brower:2000rp} (See Tab.~\ref{tb}).
In addition, we plot the curves for $q$-deformed glueball spectrum to attain further modes via $q$-deformation as can be seen from Fig.~\ref{fig1}. At $q=0.2294$ we attain $m_0=2.3699$ GeV (or 2370 MeV) that approaches the resonance X(2370) recently pointed as a possible detection of the lightest pseudoscalar glueball with spin parity $0^{-+}$ \cite{BESIII:2023wfi}. In future work, we will apply the $q$-algebra formalism to the gravity localization scenario to derive the Newtonian potential and the ``critical transitions'' with respect to the extra dimension.

\acknowledgments
We want to thank CNPq and CAPES for partial financial support. FAB acknowledges support from CNPq (Grant no. 309092/2022-1) and PRONEX/CNPq/FAPESQ-PB (Grant no. 165/2018), for partial financial support. This present article arxiv.org/abs/2403.16171 was supported by funded by SCOAP$^3$.

\section{Data Availability Statement}

The present manuscript can be found at https://arxiv.org/abs/2403.16171

\section{Declaration of interests}

The authors declare no conflict of interest.

\end{document}